\documentclass{article}
\usepackage{spconf,amsmath,graphicx,multicol,multirow,booktabs}
\usepackage[colorlinks,linkcolor=black,anchorcolor=black, citecolor=black]{hyperref}
\usepackage{amssymb}
\usepackage{enumitem}
\setlist{nosep, leftmargin=14pt}
\usepackage[numbers,sort]{natbib}
\usepackage{mwe} % to get dummy images
\title{Introducing learning rate adaptation CMA-ES into rigid 2D/3D registration for robotic navigation in spine surgery}

\name{Zhirun Zhang$^{\dag}$ \qquad Minheng Chen $^{\dag}$
\thanks{
 The two authors contribute equally to this work.
}
}
\address{$^{\dag}$School of Computer Science and Engineering, Southeast University, China 
}
\begin{document}
%\ninept
%
\maketitle
\begin{abstract}
The covariance matrix adaptive evolution strategy (CMA-ES) has been widely used in the field of 2D/3D registration in recent years. This optimization method exhibits exceptional robustness and usability for complex surgical scenarios.
%, showcasing insensitivity to local minima and troublesome noise.
However, due to the inherent ill-posed nature of the 2D/3D registration task and the presence  of numerous local minima in the landscape of similarity measures. Evolution strategies often require a larger population size in each generation in each generation to ensure the stability of registration and the globality and effectiveness of search, which makes the entire process computationally expensive.
In this paper, we build a 2D/3D registration framework based on a learning rate adaptation CMA-ES manner. The framework employs a fixed and small population size, leading to minimized runtime and optimal utilization of computing resources.
We conduct experimental comparisons between the proposed framework and other intensity-based baselines using a substantial volume of synthetic data.
The results suggests that our method demonstrates superiority in both registration accuracy and running time.
Code is available at \textit{github.com/m1nhengChen/CMAES-reg}.
\end{abstract}
\section{Introduction}
\label{sec:intro}
Intraoperative 2D/3D registration is a process aimed at aligning intraoperative 2D images, such as X-ray images, with corresponding preoperative CT scans. It is a crucial step in providing surgical planning guidance and navigation positioning for spine surgeries like percutaneous vertebroplasty and pedicle screw internal fixation. 
% This registration process provides real-time navigation guidance, allowing surgeons to accurately position and navigate during surgery, improving surgical accuracy and safety.
By registering the patient's anatomy with previously acquired high-resolution 3D images, surgeons can visualize the patient's anatomy in real-time during surgery, including the position and orientation of structures such as bones, nerves, blood vessels, implants and surgical instruments during spinal surgeries, aiding them in performing precise operations and improving surgical accuracy and safety. 

In the existing literature, intensity-based 2D/3D registration methods~\cite{chen2024optimization,op2patch} have received significant attention, and some benchmark methods for pose estimation of bone anatomies and surgical devices have been tested to meet clinical requirements in various orthopedic applications.
And in these conventional intensity-based registration methods~\cite{markelj2012review}, diverse ray-tracing techniques are employed to produce simulated two-dimensional X-ray images, commonly referred to as digitally reconstructed radiographs (DRRs), from the 3D CT volume. 
These techniques simulate the attenuation of X-rays within the human body, providing a vital component for accurate registration. 
The similarity between DRRs and X-rays is then evaluated by using some statistical-based similarity measures, \textit{ie.}, normalized cross-correlation (NCC) and mutual information (MI) or local feature representations. 
Gradient-free optimization techniques, such as Powell-Brent~\cite{powell1964efficient} and CMA-ES~\cite{hansen2001completely}, are adopted to explore the similarity function and identify both the minimum value and the corresponding patient pose within the solution space.

CMA-ES is widely regarded as one of the most prominent optimization strategies in 2D/3D registration. 
However, existing CMA-ES method still has several shortcomings. 
Specifically, in order to ensure the stability of the evolutionary path and exhaustion of the search, the original CMA-ES algorithm requires a larger population size in each generation. 
This makes the entire optimization process computationally expensive and time-consuming.
Population size adaptation mechanism~\cite{nishida2018psa} has been proposed to alleviate this drawback, but this strategy can lead to complex scheduling problems and difficulty in fully utilizing available resources.
In addition, it is worth noting that  although many learning-based 2D/3D methods~\cite{chen2023embeddedsopi} have become mainstream in recent years, intensity-based approaches have not been entirely disregarded.
They are still used to refine the prediction results of learning-based methods in many recent articles~\cite{gao2023fully,chen2024fully}.
This also implies that robust and efficient intensity-based 2D/3D registration methods are still indispensable for  automatic image-guided surgical navigation.
% methods~\cite{gao2023fully,chen2024fully} have been proposed, optimization-based methods have not been abandoned and are still used to adjust the prediction results of learning-based methods.

In this paper, we build a 2D/3D registration framework based on a learning rate adaptation CMA-ES manner. 
The framework has a fixed and small population size, which results in short running time and is well suited to achieve maximum utilization of computing resources.
We conduct experimental comparisons between the proposed framework and other intensity-based baselines using a substantial volume of synthetic data.
The results suggests that our method demonstrates superiority in both registration accuracy and running time.
\section{method}
\subsection{Preliminaries}
\noindent\textbf{Covariance matrix adaptive evolution strategy (CMA-ES).}
CMA-ES uses multivariate normal distribution to generate candidate solutions to minimize the objective function $f : \mathbb{R}^{d}\rightarrow \mathbb{R}$, 
and this distribution $N$ is parameterized by three elements: the mean vector $m \in \mathbb{R}^{d}$, the step-size $\sigma \in \mathbb{R}$, and the covariance matrix $\pi \in \mathbb{R}^{d \times d}$. 
In the iteration $t+1$, first independently sample $\lambda$ times according to the current distribution $N(m^{(t)}, \sigma^{(t)2}\pi^{(t)})$ to obtain the candidate solution $x^{(t)}_{i}$ and its corresponding value $y^{(t)}_{i}=f(x^{(t)}_{i})$on the objective function.
Next, the evolutionary path of the distribution is calculated based on the candidate solutions of the current generation obtained through sampling, and finally the distribution parameters are updated to obtain the evolved distribution $N(m^{(t+1)}, \sigma^{(t+1)2}\pi^{(t+1)})$ .
This complex update process can be simplified as follows:
\begin{equation}\label{eq1}
m^{(t+1)}=m^{(t)}+\Delta^{(t)}_m
\end{equation}
\vspace{-0.5cm}
\begin{equation}\label{eq2}
\Sigma^{(t+1)}=\Sigma^{(t)}+\Delta^{(t)}_\Sigma
\end{equation}
where $\Sigma^{(t)}$ is the verctorized representation of $\sigma^{(t)2}\pi^{(t)}$.

\noindent\textbf{Problem formulation.} The problem of rigid 2D/3d registration can be viewed as optimizing the following formula:
\begin{equation}\label{eq3}
\theta  = arg\mathop{min}_{\bar{\theta}}\mathcal{S}(\mathit{I},\mathit{P}(\bar{\theta};\textit{V}))
\end{equation}
where $I$ is the 2D fixed image, $\mathcal{S}(\cdot, \cdot)$ is the similarity function which is also the target object function $f$ and $\textit{V}$ is the 3D volume. 
The pose $\theta$ that needs to be estimated is a vector with six degrees of freedom (6DoF), $\theta = (r_x, r_y, r_z, t_x, t_y, t_z)$.
$\mathit{P}(\cdot ; \cdot)$ is the projection operator which use $\textit{V}$ and pose $\bar{\theta}$ to generate DRR.
\subsection{Learning Rate Adaptation CMA-ES}
Learning Rate Adaptation CMA-ES (LRA-CMA)~\cite{nomura2023cma} is a recently proposed variation of classic CMA-ES algorithm. 
It maintains a constant signal-to-noise ratio by adopting a learning rate adaptation mechanism. 
It is worth noting that the population number of each generation of this method is a constant default value $\lambda = 4 + \lfloor ln(d) \rfloor$, thus greatly saving computing resources. 
In simple terms, it first follows the existing CMA-ES parameter update strategy, and then calculates the learning rate factors $\delta^{(t)}_m$ and $\delta^{(t)}_\Sigma$, thereby modifying the update strategies in Eq.~\ref{eq1} and~\ref{eq2} to $m^{(t+1)}=m^{(t)}+\delta^{(t)}_m\Delta^{(t)}_m$ and $\Sigma^{(t+1)}=\Sigma^{(t)}+\delta^{(t)}_\Sigma\Delta^{(t)}_\Sigma$ respectively.

The core step of the learning rate adaptation mechanism is to adjust $\delta$ so that the signal-to-noise ratio $\eta = \alpha\delta$, where $\alpha$ is a hyperparameter. 
We first use the parameter update strategy of CMA-ES to obtain $\Delta^{(t)}_m$ and $\Delta^{(t)}_\Sigma$, then use them to calculate the new evolutionary path and current signal-to-noise ratios $\eta^{(t)}_m$ and $\eta^{(t)}_{\Sigma}$, and finally adjust the learning factor $\delta^{(t)}\leftrightarrow\delta^{(t)}_{m}/\delta^{(t)}_{\Sigma}$ accordingly to compute the updated parameters $m^{(t+1)}$, $\sigma^{(t+1)}$ and $\pi^{(t+1)}$.
% and finally update the learning factor according to the signal-to-noise ratio and calculate the final updated parameters $m^{(t+1)}$, $\sigma^{(t+1)}$ and $\pi^{(t+1)}$.
The update process of the learning rate is:
\begin{equation}\label{eq4}
\delta^{(t+1)}=\delta^{(t)}\cdot exp \bigl( min(\gamma\delta^{(t)}, \beta)\Pi_{[-1,1]}(\frac{\eta^{(t)}}{\alpha\delta^{(t)}}-1)\bigr)
\end{equation}
where $\alpha$, $\beta$ and $\gamma \in \mathbb{R}$ are hyperparameters.
In addition, in order to ensure that the optimal $\sigma$ is maintained after $\eta_m$ changes, a step-size correction strategy is adopted as shown in Eq.~\ref{eq5}.
\begin{equation}\label{eq5}
\sigma^{(t+1)}=\frac{\delta^{(t)}_m}{\delta^{(t+1)}_m}\sigma^{(t+1)}
\end{equation}
\vspace{-0.6cm}
\subsection{Application to 2D/3D Registration}
In CMA-ES-based 2D/3D method, the similarity function is regarded as the optimization objective function $f$. $m$ is the 6DoF initial patient pose $\theta^{0}$. 
In iteration $t$, we first sampled $n$ times in the solution space, \textit{i.e.} $\bar{\theta^{t}_0}, \bar{\theta^{t}_1}, \bar{\theta^{t}_2}$......$\bar{\theta^{t}_n}$, and the corresponding similarity scores $f(\bar{\theta^{t}_0}),f(\bar{\theta^{t}_1}), f(\bar{\theta^{t}_2})$......$f(\bar{\theta^{t}_n})$ are calculated, and then the evolutionary estimated pose $\theta^{t}$is obtained through the LRA-CMA optimizer.
Normalized cross-correlation (NCC) is the most common metric used to evaluate the similarity of two images $I_1$ and $I_2$ in mono-modality registration:
\begin{equation}\label{eq6}
NCC(I_1,I_2)=\sum_{i=0}^m\sum_{j=0}^n\frac{(I_1(i,j)-\bar{I}_1)(I_2(i,j)-\bar{I}_2)}{\sigma_{I_1}\sigma_{I_2}}
\end{equation}
where $\bar{I}$ and $\sigma_I$ is the pixel-wise mean and standard deviation of an $m \times n$ image. 
This method of calculating NCC directly on the entire image is also called global NCC.
In comparison, the method called local NCC (LNCC), which calculates the NCC on the local image region has a sharper similarity curve and is more difficult to converge, but it can achieve higher registration accuracy.
We use multi-scale normalized cross-correlation (mNCC)~\cite{abumoussa2023machine} as the similarity function, which is an integration of global NCC and patch-based NCC, and its landscape is smoother than gradient correlation~\cite{op2patch}.
Calculating the mNCC between the fixed image $I$ and the projected image $I_m = \mathit{P}(\bar{\theta}^{t}_i;\textit{V})$ can be expressed as:
$mNCC(I,I_m) = (1-\mu)NCC(I,I_m)+ \mu\sum_{(p_i,p_j)\in\Omega_K}LNCC(I,I_m,p_i,p_j,r)$, where NCC is the normalized cross-correlation, LNCC is the patch-based NCC and $\mu$ is a hyperparameter from (0, 1). 
The patches we use are squares with radius $r$ and center point $(p_i,p_j)$. 
In the experiment, we set $r$ to 6 and spilt the entire image evenly into $k$ non-overlapping patches.
% \begin{equation}\label{eq4}
% NCC(I,I_m)=\sum_{i=0}^m\sum_{j=0}^n\frac{(I(i,j)-\bar{I})(I_m(i,j)-\bar{I}_m)}{\sigma_{I}\sigma_{I_m}}
% \end{equation}
% \begin{tiny}
% \begin{equation}\label{eq5}
% LNCC(I,I_m,p_x,p_y,r)=\sum_{i=p_x-r}^{p_x+r}\sum_{j=p_y-r}^{p_y+r}\frac{(I(i,j)-\bar{I})(I_m(i,j)-\bar{I}_m)}{\sigma_{I}\sigma_{I_m}(2r+1)^2}
% \end{equation}
% \end{tiny}

\section{experiment}
% We evaluate our method on simulated X-ray images 
% % real X-ray data and 
% for a challenging single-view 2D/3D lumbar spine registration scenario.
\subsection{Experiment Settings}

% \noindent\textbf{Simulation study.} 
\noindent\textbf{Dataset.} The dataset consists of 52 CT scans from VerSe~\cite{sekuboyina2021verse}.  We resample the CT images to isotropic spacing of 1.0 mm and crop or pad evenly along each dimension to obtain  $256\times256\times256$ volumes with the spine ROI approximately in the center. 
We select 5 scans for hyperparameters tuning and 47 scans are used for testing. 
We define the intrinsic parameter of the X-ray simulation environment as a Perlove PLX118F C-Arm, which  has image dimensions of $1024 \times 1024$, isotropic pixel spacing of 0.199 mm/pixel, and a source-to-detector distance of 1012 mm. 
The images are downsampled to have dimensions of $256\times256$ with a pixel spacing of 0.798 mm/pixel. 

For testing, we use 1000 simulated X-ray images with angles of $U$(-20, 20) degrees in three directions, with translation in mm of $U$(-30, 30) for in-plane (X and Y) direction and $U$(-50, 50) for depth (Z) direction. The volume $V$ we use is segmented through the mask provided by the dataset to reduce the impact of soft tissue on image quality of DRRs.

\noindent\textbf{Evaluation metrics.} 
Following the standard in 2D/3D registration, we report mean target registration error (mTRE) and the error between estimated pose $\theta$ and ground truth $\hat{\theta}$ in rotation and translation respectively for our experiments.
Mean target registration error metric computes the average distance between corresponding landmarks under the estimated and ground truth poses.
Suppose we have a three-dimensional point set $\Phi$ consisting of $L$ anatomical landmarks, mTRE can be represented as:
\begin{equation}\label{eq7}
mTRE(\theta,\hat{\theta})=\frac{1}{L}\sum_{v_i\in\Phi}^L\Vert\theta\circ v_i-\hat{\theta}\circ v_i\Vert_2
\end{equation}
Besides, running time of the registration is also reported.

\noindent\textbf{Implementation details.}
The LRA-CMA optimizer is implemented through an open-source library~\cite{nomura2024cmaes}.
After hyperparameter tuning based on grid search, we set the number of generations of the evolutionary strategy to 50 and the step size $\sigma$ to 10. Our experiments were all conducted on a PC with a NVIDIA GeForce RTX 3090 GPU and  a 2.3-GHz quad-core Intel Core i7 processor.

\noindent\textbf{Baseline methods.}
We compare the proposed framework with our initial work~\cite{chen2024optimization} (CMA-ES), a well-tuned CMA-ES-based 2D/3D registration method. 
To ensure fairness, the entire experiment was conducted in a single-resolution scenario and the projection renderers of all baselines were the same as the implementation in~\cite{gao2023fully}.
In addition, a fully differentiable optimization-based method~\cite{abumoussa2023machine} (GD) is also used as a benchmark. 
Different from the original paper, we use a SGD optimizer with the momentum of 0.60 and dampening of 0.45 in PyTorch. The learning rate of the rotation components is set to 5e-2, and the learning rate for translation components is 1e-2, and a weight decay with a value of 1e-8 is adopted. 
To ensure fast convergence of the optimization, a StepLR scheduler with a step size of 25 and a gamma value of 0.9 is integrated in the framework.
We set the termination condition of the optimization as: The optimizer has searched for 200 iterations  or the minimum value of the similarity function searched within the last 100 iterations has not been updated.
% During training iteration i, we randomly sample a pair of pose parameters ($\theta^i$, $\theta^i_t$) with rotations from a normal distribution $N$(0, 10)  in degrees for all three axes, and translations $t_x,t_y,t_z$ from normal distributions $N$(0, 30), $N$(0, 15) and $N$(0, 15)  in millimeters. Additionally, we randomly select a CT volume denoted as $\textit{V}$ and its corresponding segment $V_{seg}$. The target image is generated in real-time using 
% $\textit{V}$ and $\theta^i_t$. $V_{seg}$ and $\theta^i$ are used as input to our network. The proposed framework was
% implemented with PyTorch on NVIDIA GeForce RTX 3090 GPUs with 24 GB memory. The model wass trained by using a SGD optimizer with a cyclic learning
% rate between 10 e-6 and 10 e-4 every 100 steps and
% a momentum of 0.9 for 200 k iterations. 
\vspace{-0.4cm}
\begin{table}[h!]
% \footnotesize
\begin{center}
	\label{table1}
 
	\caption{2D/3D registration performance comparing with the baseline methods.The evaluation includes measuring the mean, standard deviation and median of the mean target registration error (mTRE), as well as the pose error and registration time.}
 \setlength{\tabcolsep}{0.2mm}{
	\begin{tabular}{c|cc|cc|c} 
  \toprule[1.2pt]
  \multirow{2}{*}{Method}
  & \multicolumn{2}{|c|}{mTRE(mm)$\downarrow$}& \multicolumn{2}{|c|}{Pose error$\downarrow$} &Reg.
		\\
   \cline{2-5}    & mean(std)  & median & Rot.($^{\circ}$)& Trans.(mm)&time
  \\
\hline
  Initial & 259.2(116.8)  & 250.5&33.3&60.1& N/A \\
  CMA-ES & 171.8(87.4)& 168.0& 24.1 &72.7& 50.6 \\

GD &202.5(134.4)  & 175.0& 26.9&49.3& 28.3\\
\hline
LRA-CMA(gc) &186.8(105.7) &171.5  & 25.9&54.9& 20.7\\

LRA-CMA &169.4(92.6)&158.6 & 24.4 &55.9& 20.5\\
\bottomrule[1.2pt]
	\end{tabular}}
\label{result}	
\end{center}
\end{table}
\vspace{-0.6cm}
\begin{figure}[htb]

\begin{minipage}[b]{1.0\linewidth}
  \centering
\vspace{-0.2cm}
  \centerline{\includegraphics[width=\linewidth]{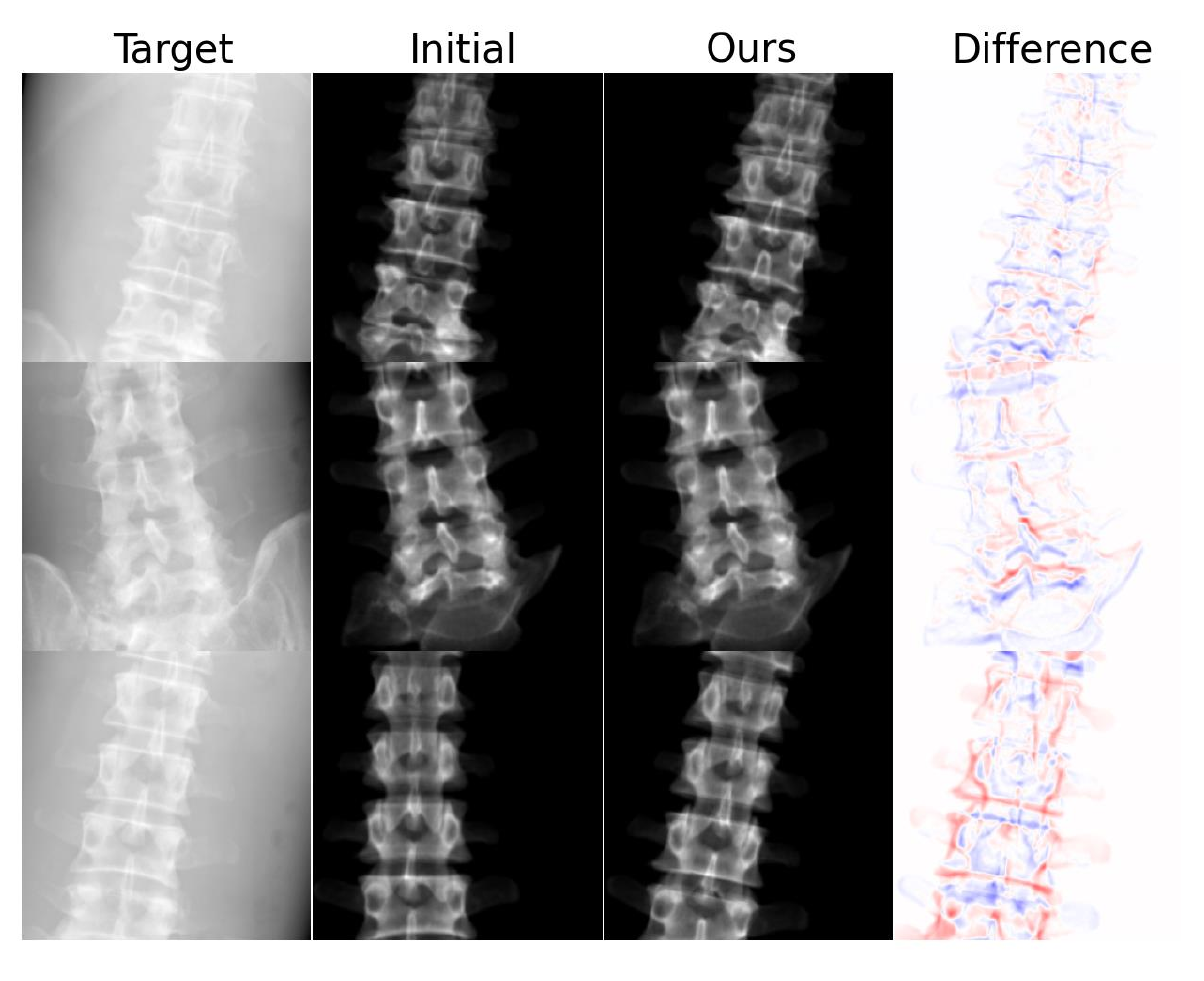}}
  \vspace{-0.6cm}
  %\centerline{(a) Result 1}\medskip
\end{minipage}
%
% \begin{minipage}[b]{.48\linewidth}
%   \centering
%   \centerline{\includegraphics[width=4.0cm]{example-image}}
% %  \vspace{1.5cm}
%   \centerline{(b) Results 3}\medskip
% \end{minipage}
% \hfill
% \begin{minipage}[b]{0.48\linewidth}
%   \centering
%   \centerline{\includegraphics[width=4.0cm]{example-image}}
% %  \vspace{1.5cm}
%   \centerline{(c) Result 4}\medskip
% \end{minipage}
%
\caption{Quantitative results of the proposed framework and the baselines. In column 1 is the preprocessed target images; in column 2 and 3 are the DRRs corresponding to the initial pose and the registration pose of our method respectively; the difference maps between the estimated poses and the ground truth are shown in column 4.}
\label{fig:result}
\end{figure}
% \begin{table}[h!]
% % \footnotesize
% \begin{center}
% 	\label{table1}
 
% 	\caption{2D/3D registration performance comparing with the baseline methods. This evaluation includes measurement of the mean Target Registration Error (mTRE) at the 50th, 75th, and 95th percentiles, as well as calculating the success rate (SR) of registration.}
%  \setlength{\tabcolsep}{3.5mm}{
% 	\begin{tabular}{c|ccc|c} 
%   \toprule[1.2pt]
%   \multirow{2}{*}{Method}
%   & \multicolumn{3}{|c|}{mTRE(mm)$\downarrow$}& \multirow{2}{*}{SR(\%)$\uparrow$} 
% 		\\
%    \cline{2-4}    & 95th & 75th & 50th & 
%   \\
% \hline
%   Initial & 225.7 &188.6 & 148.0&    \\
%   +CMA-ES & 98.6&55.7 & 24.2& 22.0  \\
% \hline
% ProST &185.7 &151.2 &114.3&  \\
%   +CMA-ES &38.3 &12.8 &2.6 &  55.6\\
% \hline
% SOPI &163.6&133.1 &101.3 &  \\
%   +CMA-ES &34.7&9.1 &2.2 & 58.4 \\
% \hline
% \textbf{Ours} & 155.7& 127.2& 95.1&  \\
%   +CMA-ES &32.0&7.9&2.2 & \textbf{61.0} \\
% \bottomrule[1.2pt]
% 	\end{tabular}}
% \label{result}	
% \end{center}
% \end{table}
\vspace{-0.6cm}
\subsection{Results}
During experiment, we randomly sample poses with rotations from a normal distribution $N$(0, 10) in degrees for all three axes, and translations $t_x,t_y,t_z$ from normal distributions $N$(0, 15) in millimeters as the initial value. 
As shown in Table~\ref{result}, We compare our method with two intensity-based baselines.
And we also compared the performance of different similarity functions in 2D/3D registration, specifically GC vs. mNCC.
Because the experiment is a single-view and low-resolution registration scene, and considering the impact of image noise and the intensity-based method inherently only has a smaller capture range. So there is a visible error between the results of all our experiments and the ground truth.
The GD method  holds significant potential, but its current challenge lies in the exhaustive resource demand for pre-grid search of hyperparameters , along with its limited generalization capability.
The LRA-CMA method that uses mNCC as the similarity function performs better than the baseline using GC, which reflects that this metric is more effective in registration scenarios with larger offsets.
Compared with the CMA-ES method, although the proposed method does not have a significant improvement in accuracy, the LRA-CMA method runs over twice faster than the CMA-ES benchmark. 
This shows that the use of an evolutionary strategy with a learning rate adaptive mechanism and a default population size can indeed solve the complex 6DoF rigid 2D/3D registration problem.
Moreover, we provide several qualitative examples of the proposed  method in Fig.~\ref{fig:result}.
\vspace{-0.3cm}
\section{conclusion}
In this work, we introduce the learning rate adaptation covariance matrix adaptive evolution strategy into the field of rigid 2D/3D registration. 
By comparing with other existing intensity-based benchmarks, the registration method using LRA-CMA as the optimizer has less running time and higher registration accuracy.
We also note the potential of this algorithm in resource utilization. Future work will focus on parallel GPU acceleration of this algorithm to maximize the use of computing resources and reduce the running time of the framework,  as well as adjusting the current hyperparameter settings and parameter selection strategies to further improve the registration performance of the framework.
% \vspace{-0.3cm}
% \section{Compliance with Ethical Standards}
% % % This study was performed in line with the principles of the Declaration of Helsinki. Approval was granted by the Ethics Committee of University B (Date…/No. …).
% % This research study was conducted using human subject data.
% % The institutional review board at the local institution approved
% % the acquisition of the data, and written consent was obtained
% % from the subject.
% This research study was conducted retrospectively using human subject data made available in open access at github.com /anjany/verse under the CC BY-SA 4.0 License. Ethical approval was not required as confirmed by the license attached with the open access data.
\vspace{-0.3cm}
\section{acknowledgement}
This work was supported in part by Jiangsu provincial joint international research laboratory of medical information processing.  
\bibliographystyle{ieeetr}
\bibliography{strings,refs}

\end{document}